\documentclass[aps,pra,floatfix,twocolumn,superscriptaddress]{revtex4-2}

\usepackage{mathtools}

\usepackage{amsmath}
\usepackage{amsfonts}
\usepackage{amssymb}
\usepackage{graphicx}
\usepackage{parskip}
\usepackage{xcolor}

\long\def\comment#1{}

\begin{document}

\title{Intermediate-temperature topological  Uhlmann phase on IBM quantum computers}

\author{Christopher Mastandrea}
\affiliation{Department of Physics, University of California, Merced, CA 95343, USA}

\author{Costin Iancu}
\affiliation{Lawrence Berkeley National Laboratory, Berkeley, CA 94720, USA}

\author{Hao Guo}
\affiliation{School of Physics, Southeast University, Jiulonghu Campus, Nanjing 211189, China}
\affiliation{Hefei National Laboratory, Hefei 230088, China}

\author{Chih-Chun Chien}
\email{cchien5@ucmerced.edu}
\affiliation{Department of Physics, University of California, Merced, CA 95343, USA}

\begin{abstract}
A spin-1 system can exhibit an intermediate-temperature topological regime with a quantized Uhlmann phase sandwiched by topologically trivial low- and high-temperature regimes. We present a quantum circuit consisting of system and ancilla qubits plus a probe qubit which prepares an initial state corresponding to the purified state of a spin-1 system at finite temperature, evolves the system according to the Uhlmann process, and measures the Uhlmann phase via expectation values of the probe qubit. Although classical simulations suggest the quantized Uhlmann phase is observable on IBM's noisy intermediate-scale quantum (NISQ) computers, an implementation of the circuit without any optimization exceeds the gate count for the error budget and results in unresolved signals. Through a series of optimization with Qiskit and BQSQit, the gate count can be substantially reduced, making the jumps of the Uhlmann phase more visible. A recent hardware upgrade of IBM quantum computers further improves the signals and leads to a clearer demonstration of interesting finite-temperature topological phenomena on NISQ hardware.
\end{abstract}

\maketitle

\section{Introduction}
Discoveries of topological properties behind physical systems have revolutionized the classification of materials and matter~\cite{KaneRMP,ZhangSCRMP,ChiuRMP}. The Berry phase of pure states~\cite{Berry84,Nakahara}, which is the holonomy of the underlying fiber bundle of pure states, lays the foundation for understanding many topological concepts. A generalization of the Berry phase to mixed states based on the Uhlmann bundle of mixed states~\cite{Uhlmann86,Uhlmann89,Uhlmann95} provides a natural reflection of the holonomy of statistical ensembles described by density matrices. More recent studies of the Uhlmann phase have shown promising results of using it as a finite-temperature topological indicator~\cite{Viyuela14,PhysRevLett.113.076407,OurPRA21,OurUB,Galindo21,Zhang21,Wang25}. While physical implications of the Uhlmann phase were not clear in the early days, Ref.~\cite{Viyuela2018} has demonstrated the simulation and measurement of the Uhlmann phase of a two-level system on the IBM quantum computers. As the mixedness of the system increases, the Uhlmann phase jumps from $\pi$ to $0$ abruptly, showing a  finite-temperature topological phase transition.

Finite-temperature quantum systems in equilibrium are also described by mixed states. Contrary to the common belief that temperature tends to destroy topological properties, a finite-temperature topological regime characterized by finite values of the Uhlmann phase sandwiched by trivial low- and high-temperature regimes with vanishing Uhlmann phase has been found in spin-1 systems with three levels~\cite{OurPRA21}. The topological changes at finite temperatures originate from the twisting of the parallel transport in the Uhlmann bundle as the system traverses a loop in the parameter space. In the following, we implement a similar quantum circuit of Ref.~\cite{Viyuela2018} to simulate a spin-1 (three-level) system coupled to an ancilla modeling the environment going through the Uhlmann process and extract the temperature-dependent Uhlmann phase on the IBM quantum computers. To overcome the limitation that quantum computers take state vectors and operates unitary transformations, we first use purification of density matrices to rewrite them as purified states, which can be represented by state vectors. The Uhlmann process compatible with the Uhlmann parallel-transport condition is translated into a series of time-evolution operators for the system and ancilla, so the topological Uhlmann phase can be accumulated at the end of the process.

Extracting the Uhlmann phase at the end of the quantum circuit, however, has been a challenge until Ref.~\cite{Viyuela2018} shows a scheme by coupling the system and ancilla to an additional probe qubit and extract the phase from its expectation values. The reason why adding a single qubit suffices to extract the Uhlmann phase is because such a procedure is actually a generalization of a scheme known as the deterministic quantum computation with one qubit (DQC1)~\cite{Knill98}, which extracts the expectation value of a mixed state going through a unitary process. The DQC1 has been realized in early experiments~\cite{Lanyon08,Passante11,Hor15} and may have applications in quantum machine learning~\cite{Kim25}. While the original DQC1 may be considered as a generalization of the Hadamard test of pure states to the infinite-temperature density matrices proportional to the identity matrix, a replacement by finite-temperature density matrices from the canonical ensemble allows the scheme to extract the relevant expectation values to obtain the Uhlmann phase. While Ref.~\cite{Viyuela2018} uses density matrices with a parameter to alter the mixedness, here we will present the results of thermal states with temperature effects on the IBM quantum computers by encoding the temperature into the probability amplitudes of the corresponding purified states.

Currently available quantum computers are of the noisy intermediate-scale quantum (NISQ) type with non-negligible intrinsic error rates~\cite{Preskill2018quantumcomputingin,RevModPhys.94.015004}. Our studies of the quantized Uhlmann phase at finite temperatures serve the following purposes. First, the quantized values of the Uhlmann phase naturally provides intrinsic immunization against small errors and fluctuations. Differentiating the Uhlmann phase thus provides a class of physical problems which can be handled by NISQ hardware. Second, the Uhlmann phase of spin-$j$ systems have been analyzed in details~\cite{OurPRA21,Galindo21}. By increasing the value of $j$, the Uhlmann-phase simulation provides a benchmark for NISQ computers and their optimization software. We discuss synthesis based compilation workflows for resource optimization as deployed by the Qiskit and BQSkit~\cite{osti_1785933} compilation infrastructures.  Third, the quantum computer itself may be considered as an experimental apparatus for measuring the elusive Uhlmann phase. As pointed out in Ref.~\cite{OurDPUP}, the Uhlmann process is typically incompatible with Hamiltonian dynamics, thereby making it challenging to probe the genuine Uhlmann phase in natural or artificial materials in the lab. Nevertheless, quantum computers allow for precise controls of both the system and ancilla to follow the Uhlmann parallel-transport condition, thereby providing direct probes of the Uhlmann phase. For the spin-1 system analyzed here, the finite-temperature topological regime offers valuable examples of the complex interplay between topology and temperature of quantum systems. 

The rest of the paper is organized as follows. Sec.~\ref{sec:theory} briefly introduces the Uhlmann phase and the quantum circuit for realizing and measuring it on quantum computers. Sec.~\ref{sec:spin1} shows the implementation of a spin-1 system on IBM's quantum computers. Sec.~\ref{sec:result} shows the results from the quantum circuit without optimization and compare them with those with different levels of optimization and hardware upgrade. The intermediate-temperature topological regime on NISQ computers can be seen clearly after the optimization and upgrade. Sec.~\ref{sec:discussion} discusses the statistical distance in the state preparation, BQSKit approximation, error mitigation, and other subtleties and applications of the circuit and optimization. Finally, Sec.~\ref{sec:conclusion} concludes our work. The Appendix presents the results of the Uhlmann phase of a spin-$1/2$ system as a comparison. 

\section{Theory and Circuit}\label{sec:theory}

\subsection{Uhlmann phase}
We summarize the Uhlmann phase, focusing on spin-$j$ systems, following Ref.~\cite{OurPRA21}. The Uhlmann phase of mixed states is constructed through purification of the density matrices via $\rho = WW^{\dagger}$, where $W$ is called the purification or amplitude of the density matrix. For full-rank density matrices including thermal states at finite temperatures, the purification can be uniquely determined as $W = \sqrt{\rho}U$ where $U$ is a unitary matrix that will contain the phase information, analogous to the phase factor $\exp(-i\theta)$ in the pure-state description. These amplitudes have a one-to-one correspondence with the purifed states $|W\rangle$ and introduce a Hilbert space $\mathcal{H}_{W}$ in which the inner-product is given by the Hilbert-Schmidt product $\langle W_{1}|W_{2} \rangle = Tr(W_{1}^{\dagger}W_{2})$.

Two purifications are said to be parallel to each other if $W_{1}^{\dagger}W_{2} = W_{2}^{\dagger}W_{1} > 0$. The parallel condition is not transitive, meaning that the final purification may no longer be "in phase" with the initial purification. By requiring maximal parallelity between adjacent states in a process, the Uhlmann parallel transport condition is given by $\dot{W}^{\dagger}W = W^{\dagger} \dot{W}$, where $\dot{W} = \frac{dW}{dt}$. When the system traverses a closed loop in the parameter space, the difference between the initial and final purifications following the Uhlmann parallel-transport condition corresponds to the Uhlmann holonomy. The Uhlmann phase is a scalar representation of the Uhlmann holonomy.

Explicitly, the Uhlmann process assumes the system is governed by a set of parameters 
spanning the parameter space.  For a cyclic process with $\rho(t = 0) = \rho(t=\tau)$, where $t$ is a parametrization of the curve in the parameter space with total parametrized length $\tau$. The Uhlmann parallel-transport condition rules out accumulation of the dynamic phase, thereby rendering the Uhlmann phase topological. However, the requirement of the Uhlmann parallel-transport condition needs coordinated evolution of the purified state since the Uhlmann process is not compatible with simple Hamiltonian dynamics~\cite{OurDPUP}. Nevertheless, the evolution operators for the Uhlmann process can be found for exemplary systems, including spin-$j$ systems, which allows systematic studies of the Uhlmann phase via simulations on quantum computers.

From the Uhlmann parallel-transport condition, the initial and final phase factors of the amplitude differ by the Uhlmann holonomy \cite{OurPRA21}:
\begin{equation}\label{eq:Utau}
    U(\tau) = \mathcal{P} [\exp(-\oint A_{U})] ~U(0),
\end{equation}
where $A_{U} = -dUU^{\dagger}$ is the Uhlmann connection and $\mathcal{P}$ is the path-ordering operator. For a cyclic process with the initial phase factor $U(0)$ set to be the identity operator, the overlap between the initial and final purifications is known as the Loschmidt amplitude $\mathcal{G}^{U} = \langle W(0)| W(\tau)\rangle = Tr[\rho(0)U(\tau)]$. The Uhlmann phase can then be found by
\begin{equation}\label{eq:thetaU}
\theta_{U} = arg(\mathcal{G}^{U}) = arg(Tr[\rho(0)U(\tau)]).
\end{equation}

When the initial and final purifications under consideration are orthogonal, meaning  a Loschmidt amplitude of $\mathcal{G}^{U} = 0$, a jump in the Uhlmann phase value appears, signifying the presence of a topological phase transition \cite{Viyuela14,OurPRA21}. Given the quantized jumps of the Uhlmann phase across the critical points, the measurement of the Uhlmann phase of spin-$j$ system becomes a suitable benchmark problem for currently available NISQ computers.

\subsection{Simulation and measurement of  the Uhlmann phase}
\begin{figure}[t]
\centering

\includegraphics[width=\columnwidth]{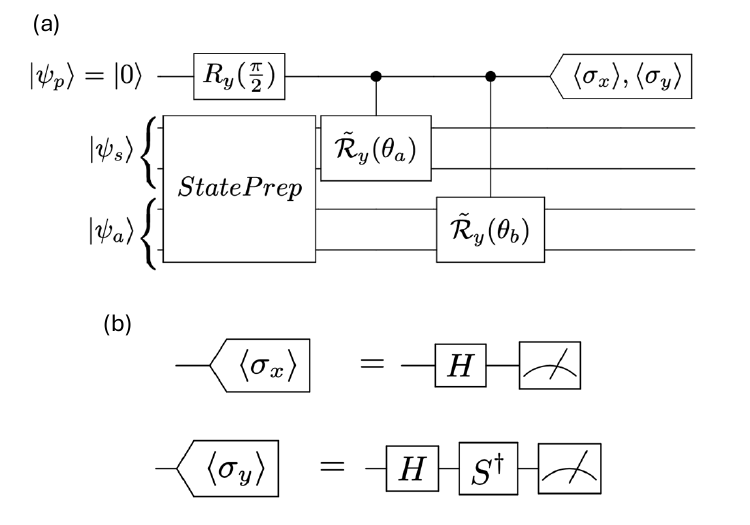}

\caption{(a) Quantum circuit for the realization of the Uhlmann phase of the spin-1 system with two qubits for the system and two qubits for the ancilla. Only the probe qubit is measured at the end of the circuit. The top-most probe qubit is measured in the computational X and Y basis in two separate circuits to preform the full tomography. The $\tilde{\mathcal{R}}_{y}$ gates represent the two-qubit Ry rotation gate. (b) Circuits for $\langle \sigma_{x} \rangle, \langle \sigma_{y} \rangle$ measurements. The full circuit is to be run over many shots to generate the expectation values.
}
\label{fig:circuit}
\end{figure}

For generating and measuring the Uhlmann phase in a spin-$j$ system, we generalize the procedure outlined in Refs~\cite{Viyuela2018,OurPRA21} and divide the process into three different blocks: State preparation, Uhlmann process, and finally the generalized DQC1 scheme for extraction of the Uhlmann phase. Fig.~\ref{fig:circuit} illustrate the process. Since a quantum computer takes and processes state vectors, we need to map density matrices to state vectors in a systematic way. The initial state of the system and ancilla qubits are prepared into the entangled state representing the purified state of the initial density matrix of the system. The initial state is in thermal equilibrium at temperature $T$ described by the canonical ensemble: 
\begin{equation}
\rho = \frac{1}{Z} e^{-\beta \hat{H}}.
\label{eq:thermalStates}
\end{equation} 
Here $\hat{H}$ is the Hamiltonian, $Z=Tr(e^{-\beta \hat{H}})$ is the partition function, and $\beta = \frac{1}{k_B T}$. In the following, we will set $k_B=1=\hbar$. We follow the purification of the density matrix of the thermal state in the diagonal form $\rho=\sum_j \lambda_j |j\rangle\langle j|$ to construct the amplitude as $W=\sum_j \sqrt{\lambda_j} |j\rangle\langle j|$, which has a correspondence with the purified state $|W\rangle=\sum_j \sqrt{\lambda_j} |j\rangle |j\rangle$. Therefore, the state preparation is to generate a superposition of states in the $|j\rangle |j\rangle$ basis with probability amplitude $\sqrt{\lambda_j}$. Here the first (second) $|j\rangle$ refers to the system (ancilla).

The generation of the Uhlmann phase $\theta_U$ from Eq.~\eqref{eq:thetaU} can be mapped to the applications of two engineered evolution operators $U_a$ and $U_s$ on the system and ancilla qubits to respect the Uhlmann parallel-transport condition, as discussed in Ref.~\cite{Viyuela2018}. Here we present the explicit forms of the evolution operators following the formalism of Ref.~\cite{OurPRA21}. 
We consider a spin-$j$ system coupled to an external magnetic field described by the Hamiltonian 
\begin{equation}\label{eq:H}
\hat{H}=\omega_0\mathbf{J}\cdot\mathbf{\hat{B}}
\end{equation}
with $\mathbf{\hat{B}}=(\sin\theta\cos\psi, \sin\theta\sin\psi,\cos\theta)$. $\omega_0$ will be the unit for $T$ in the following. When the system traverses a great circle with a constant $\psi$ in the parameter space, the evolution operators for generating the Uhlmann process are given by 
\begin{eqnarray}
    U_{s}(t) &=& \exp(-i\int_{0}^{t}\theta^{\prime} dt^{\prime}J_{y}), \nonumber \\
    U_{a}(t) &=& \exp(-i \eta \int_{0}^{t}\theta^{\prime} dt^{\prime}J_{y}).
    \label{eq:UhlmannUnitaries}
\end{eqnarray}
Here $\eta = sech(\frac{\beta \omega_{0}}{2})$ and $\theta^{\prime} = \frac{d\theta(t^\prime)}{dt^\prime}$ with $\theta(t) $ being a loop of longitude. Here $J_{y}$ is the y-component of the angular momentum operator. Consequently, these unitary operators are equivalent to rotations about about the y-axis and can thus be translated into the Ry rotation operators on the system and ancilla qubits, respectively. However, the measurement of the Uhlmann phase described below requires the Uhlmann process to be controlled by an additional probe qubit. Therefore, the $U_s$ and $U_a$ evolution operators need to be generalized to control-rotations for spin-$j$ systems in the quantum circuits.

We will adapt the modified DQC1 scheme of Ref.~\cite{Viyuela2018} to the finite-temperature spin-$j$ formalism of Ref.~\cite{OurPRA21} by coupling the mixed state with a pure probe qubit prepared in the $|+\rangle=(1/\sqrt{2})(|0\rangle+|1\rangle)$ state. The probe qubit then controls the evolution operators of both system and ancilla that generate the Uhlmann process. Estimation the Uhlmann phase is achieved through measuring certain expectation values of the reduced density matrix of the probe qubit. The combined evolution of the system (s) and ancilla (a) from the purification leads to the evolution operator $U_m$ on the mixed state (m), and the total density matrix $\rho_{pm}$ of the mixed state and probe qubit (p) becomes  
\begin{eqnarray}
\rho_{pm}=\left ( \begin{array}{cc}
\rho_{m} & \rho_{m}U_{m} \\
U_{m}^{\dagger}\rho_{m} & U_{m}^{\dagger}\rho_{m}U_{m}
\end{array}\right).
\end{eqnarray}
Here $\rho_m$ is the reduced density matrix of the mixed state consisting of the system and ancilla. A partial trace over the mixed state then gives the reduced density matrix $\rho_p$ of the probe
\begin{eqnarray}\label{eq:rho_p}
\rho_{p}&=&Tr_m(\rho_{pm}) \nonumber \\
&=&\left ( \begin{array}{cc}
1 & Tr_m(\rho_{m}U_{m}) \\
Tr_m(U_{m}^{\dagger}\rho_{m}) & 1
\end{array}\right).
\end{eqnarray}
Here $Tr_m(\rho_m)=1$ and the cyclic properties of the trace have been applied. By identifying $U_m$ with the Uhlmann process described by Eq.~\eqref{eq:Utau}, the Uhlmann phase defined in Eq.~\eqref{eq:thetaU} can then be calculated by $\theta_{U} = arg(\langle \sigma_{x} \rangle_p + i\langle \sigma_{y} \rangle_p)$. Here the subscript $p$ denotes the average with respect to the probe qubit only. In the following, we will drop the subscript $p$ if there is no confusion. Fig.~\ref{fig:circuit} shows the extractions of the relevant expectation values of the probe qubit. We mention that the system and ancilla qubits of the mixed state are left to decohere in order to fulfill the partial trace in Eq.~\eqref{eq:rho_p}.









\subsection{Uhlmann phase on IBM Qiskit platform}
We construct the quantum circuit to generate and measure the Uhlmann phase using IBM's Qiskit library \cite{qiskit2024} with the goal of testing on the cloud-based QPU hardware. 
Because of the nature of the DQC1 measurement procedure, we have two separate circuits that need to be run to measure $\langle \sigma_{x} \rangle$ and $\langle \sigma_{y} \rangle$ respectively in order to calculate the Uhlmann phase for a single temperature. However, each circuit has identical state preparation and Uhlmann process.


As outlined in Ref. \cite{OurPRA21}, a general method for preparing an arbitrary quantum state of the system and ancilla in the computational basis suffices to accomplish the state preparation for the purified state representing an initial density matrix in thermal equilibrium. Take the spin-1/2 system for example, the density matrix has the form $\rho(0)=\rho_{0}|0\rangle\langle 0|+\rho_{1} | 1\rangle\langle 1|$ with $\rho_{0,1} = \frac{\exp(-\beta E_{0,1})}{Z}$. To prepare the corresponding purified state $|W(0)\rangle=\sqrt{\rho_0}|00\rangle+\sqrt{\rho_1}|11\rangle$, a series of (controlled) rotation operators are in place. The appendix of Ref.~\cite{OurPRA21} has more details and illustrations of the arbitrary state preparation.
In general, the method introduces $2^n$ parameters for the rotation angles in order to prepare an arbitrary $n$-qubit state. Consequently, this method becomes ineffective as the number of system and ancilla qubits increase.
Moreover, the arbitrary-state preparation requires a large number of X and multi-qubit controlled rotation gates to each possible state. This contributes to increasing the circuit depth by a significant amount when converted to the native Instruction-Set-Architecture (ISA) gates used by IBM's QPUs. Later on we will introduce another more optimized method for the state preparation.


The Uhlmann process is realized by applying the system and ancilla evolution operators given in Eq.~\eqref{eq:UhlmannUnitaries} to the respective system and ancilla qubits. As mentioned, these unitary operators can be translated into rotation gates with the rotation angles determined by the Uhlmann process. For the case of a spin-1/2 system, the Uhlmann process can be constructed by using the standard gates that are included in Qiskit. When moving into higher spin systems, however, it will become necessary to construct these gates as most software packages only provide the operators and gates for two-level systems. Qiskit however, allows for the construction of a circuit to carry out the desired unitary matrix for the operation through the UnitaryGate function, which generates the circuit according a given unitary matrix. When implementing the Uhlmann process for the spin-1 system, we will construct the needed gates using this method to create the system and ancilla evolution operators which respect the Uhlmann parallel-transport condition and produce the Uhlmann phase in the mixed state represented by the system and ancilla.

After promoting the unitary operators for the Uhlmann process to controlled operators by the additional probe qubit, the Uhlmann phase is extracted by measuring the expectation values of the Pauli $\sigma_{x}, \sigma_{y}$ matrices of the probe qubit. Thus, for each temperature we run the circuit multiple times to generate the needed statistics to estimate these expectation values. Each circuit is run for 2024 shots and the $\langle \sigma_{x} \rangle$ and $\langle \sigma_{y} \rangle$ values are extracted from the measurements of the probe qubit to obtain the Uhlmann phase for one temperature encoded by the initial state. For brevity in presenting our circuits, we will define these measurements using the labels shown in Fig. \ref{fig:circuit} where it is implied that the circuits are run for many shots to create the corresponding expectation values.

\section{Uhlmann phase of a spin-1 system on quantum computers}\label{sec:spin1}
\subsection{Construction of quantum circuit}
A spin-1 system has three levels, which may be realized by a quitrit or two qubits. Here we follow the latter approach and group two qubits to represent a spin-1 object since currently available quantum computers are mostly of the qubit type. To this end then, we chose to use IBM's cloud-based system but paid the price of projecting out one degree of freedom for each pair of qubits to realize a spin-1 object. To accommodate the spin-1 system plus its spin-1 ancilla along with a probe qubit for the generalized DQC1 scheme, we will use a total of five qubits.

The addition of two spin-1/2  systems creates a spin-1 system with four states. The $|S=0,S_z=0\rangle$ state is an antisymmetric singlet state given by $\frac{1}{\sqrt{2}}(|01 \rangle - |10 \rangle)$.
The symmetric triplet states are $|S=1,S_Z=1\rangle=|00\rangle$, $|S=1,S_z=0\rangle=\frac{1}{\sqrt{2}}(|01 \rangle + |10 \rangle)$, and $|S=1,S_Z=-1\rangle=|11\rangle$.
These states make up the basis for what we will be calling the "physical basis". With $|S,S_Z\rangle$ labeled with a "ph" subscript, we have the basis $(|0, 0\rangle_{ph},|1, 1\rangle_{ph}, |1, 0\rangle_{ph}, |1, -1\rangle_{ph})^T$. 
For our work, we encode and manipulate the physical triplet states of the spin-1 system while leaving the singlet state alone. The translation between the physical basis and the two-qubit computational basis $(|00\rangle,|01\rangle,|10\rangle,|11\rangle)^T$ is facilitated by the unitary matrix 
\begin{equation}
    \mathcal{M} = 
    \begin{pmatrix}
        0 & \frac{1}{\sqrt{2}} & -\frac{1}{\sqrt{2}}  & 0 \\
        1 & 0 & 0 & 0 \\
        0 & \frac{1}{\sqrt{2}} & \frac{1}{\sqrt{2}}  & 0 \\
        0 & 0 & 0 & 1
    \end{pmatrix}.
\end{equation}
Explicitly, $|\psi \rangle_{ph} = \mathcal{M}|\phi \rangle$.
Importantly, the gate operations $U$ that we preform for the Uhlmann evolution described in Eq.~\eqref{eq:Utau} will be constructed from its action on the physical basis states. To build the quantum circuit in the computational basis, each operation will involve transforming from the computational basis to the physical basis, applying the operation, and then transforming back into the computational basis, i.e., $\mathcal{M}^{\dagger}U\mathcal{M}|\phi\rangle$.

When working with the physical basis, care must be taken when constructing unitary operators and state preparation to ensure that the anti-symmetric singlet state $|0, 0\rangle_{ph}$ is projected out in each operation. In practice, we achieve this by constructing the operation $\hat{A}$ for the triplet states and then embedding it in the unitary two-qubit gate $V$ without affecting the singlet state. This can be done by constructing the unitary $\hat{V} = 
\begin{pmatrix} 
1 &0 &0 &0\\
0& & & \\
0& & \hat{A} & \\
0 &  &  & 
\end{pmatrix}$ in the physical basis and then transforming it to the computational basis by the similar transformation facilitated by $\mathcal{M}$.
For the Uhlmann evolution shown in Fig.~\ref{fig:circuit}, controlled rotations with the probe qubit as the control and either the spin-1 system or spin-1 ancilla as the target are implemented in the computational basis.

We caution that the choice of operators in the spin-1/2 system studied in Ref.~\cite{Viyuela14} comes from two circuit identities, firstly that a controlled-Ry gate can be decomposed into two Ry gates and two CNOT gates, and secondly that we can reverse the order of these operations without affecting the outcome of the circuit \cite{Nielsen_Chuang_2010}. For the spin-1 circuit, however, these identities do not hold true in general. Consequently, we cannot reuse these operators and have to use the full controlled-Ry gate to accomplish the Uhlmann process.

When considering a controlled-Ry gate on a spin-1 system, a two-qubit Ry gate $\mathcal{R}_{y}$ is constructed in the physical basis according to the rotational operator for the triplet states. It is then translated to $\tilde{\mathcal{R}_{y}}=\mathcal{M}^{\dagger}\mathcal{R}_{y}\mathcal{M}$ in the computational basis and has the explicit form
\begin{math}
\tilde{\mathcal{R}_{y}} = \frac{1}{2}
    \begin{pmatrix}
        1 + cos(\theta) & -sin(\theta) & -sin(\theta) & 1-cos(\theta) \\
        sin(\theta) & 1+cos(\theta) & cos(\theta) -1 & -sin(\theta) \\
        sin(\theta) & cos(\theta)-1 & 1+cos(\theta) & -sin(\theta) \\
        1-cos(\theta) & sin(\theta) & sin(\theta) & 1+cos(\theta) \\ 
    \end{pmatrix},
\end{math}
where $\theta$ is the rotation angle. 
We can further construct the controlled version of the rotation gate as
\begin{math}
    C\tilde{\mathcal{R}_{y}} =     
    \begin{pmatrix}
        I_{4\times4} & O_{4\times4}\\
        O_{4\times4} & \tilde{\mathcal{R}_{y}}
    \end{pmatrix}.
\end{math}
Here $I_{4\times4}$ and $O_{4\times4}$ are the $4\times4$ identity and zero matrices, respectively.

For the spin-1 system and ancilla, the unitary operators $U_s$ and $U_a$ for generating the Uhlmann process angles are given by Eq. \eqref{eq:UhlmannUnitaries} with
\begin{math}
    J_{y} = \frac{1}{i\sqrt{2}}
    \begin{pmatrix}
        0 & 1 & 0 \\
        -1 & 0 & 1 \\
        0 & -1 & 0 \\
    \end{pmatrix}
\end{math}
in the triplet space.
By rewriting $U_s$ and $U_a$ as $\mathcal{R}_y$ with suitable angles, embedding them into the physical basis, transforming them to the computational basis, and forming the controlled rotation operators shown in Fig.~\ref{fig:circuit}, the Uhlmann process can be simulated on quantum computers with the probe qubit to extract its value.

\section{Results}\label{sec:result}
We present the extraction of the spin-1 Uhlmann phase before and after optimization on different generations of the IBM quantum computers. Before running the full circuit on the IBM QPUs, we first put it through the Aer simulator provided by Qiskit. This simulator was initialized with the built-in noise model that bases itself off of the available system calibration data for the corresponding IBM QPUs, which were IBM-Sherbooke and IBM-Kingston in our study. Explicitly, the noise model in the simulation includes effects from the last calibrated gate error rates, the $T_{1}$ and $T_{2}$ decoherence times for each qubit, the application time of each gate, and the measurement readout error rates. While including more realistic effects from the real system calibration data, the noise model might not account for error propagation from gate errors or errors from other sources \cite{Li2025}. The results from the noise model on the simulator are shown by the dotted black curves in Fig. \ref{fig:UhlmannResults} for different circuit implementations and different QPUs. Here we can see that while the simulations with built-in noise do not exactly follow the analytic results, the qualitative behavior of the intermediate-temperature topological regime characterized by jumps of the Uhlmann phase is still recognizable on all cases presented in Fig. \ref{fig:UhlmannResults}.

\subsection{Naive implementations on NISQ hardware}
\begin{figure}[t]
    \centering
    \includegraphics[width=0.85\columnwidth]{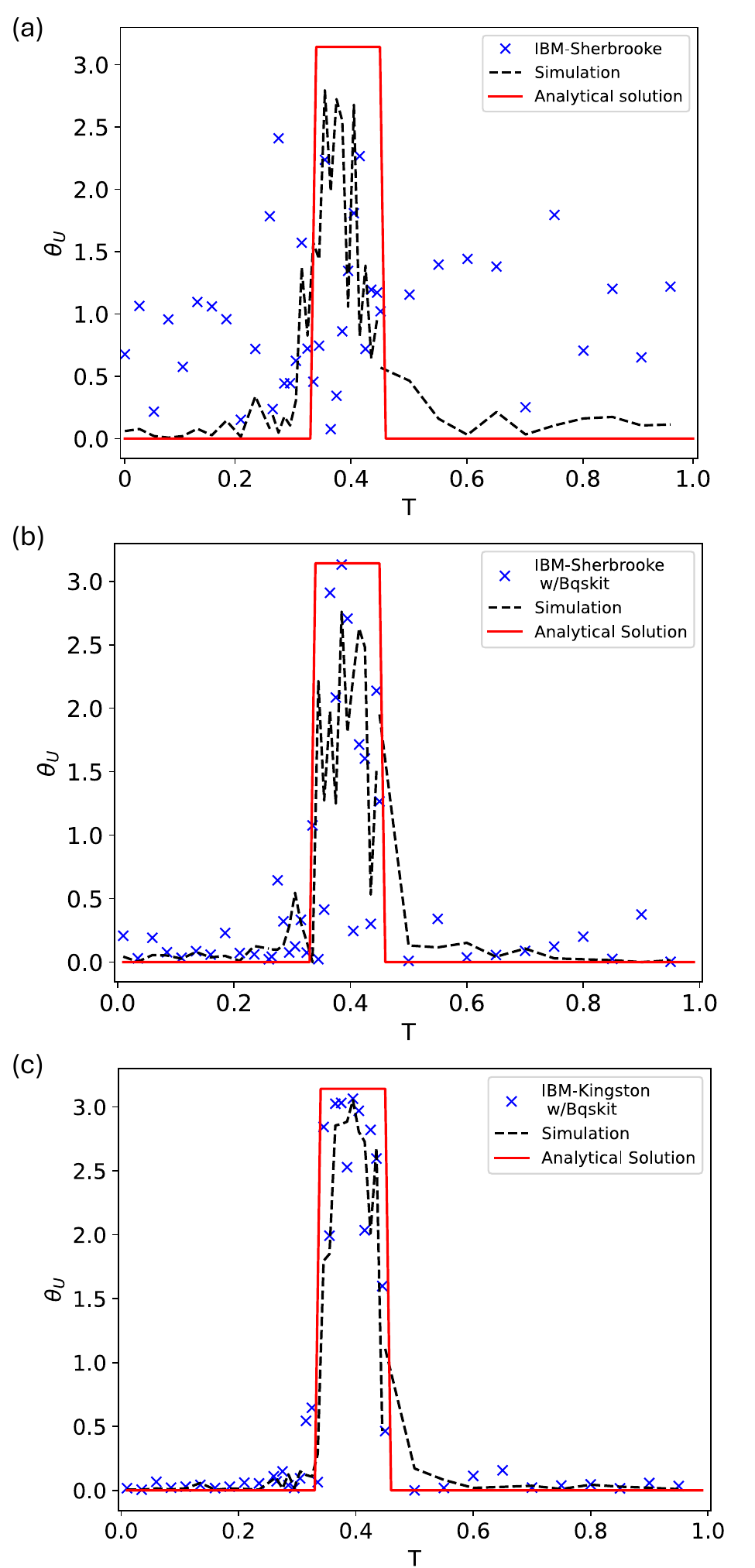}
    \caption{Analytical predictions (solid lines), noisy simulations (dashed lines), and QPU results (cross symbols) of the Uhlmann phase of the spin-1 system for the naive circuit without optimizations (a) and circuits with Qiskit state preparation and BQSKit optimizations run on IBM-Sherbrooke (b) and IBM-Kingston (c). Moving from panel (a) to (b) signifies the optimization of the circuit on the same QPU (IBM Sherbrooke), whereas moving from panel (b) to (c) signifies an upgrade from the Eagle R1 to Heron R2 QPU. For panels (b) and (c) the default approximation distance value of $\varepsilon = 10^{-8}$ was used within the BQSKit compiler.}
    \label{fig:UhlmannResults}
\end{figure}

When running the circuit without any optimization on actual hardware, however, we find that these measured results differs greatly from the expected results seen in the simulator, as shown in Fig.~\ref{fig:UhlmannResults}(a). The deviations of the results are so significant that even the $\pi$ jump of the Uhlmann phase from the QPU cannot be resolved. Upon a closer examination of the circuit, we saw the gate counts reach about 2000 for state preparation and about 300 for the Uhlmann process. Therefore, the circuit without optimization exceeds the error budget of the QPU and compromises the resolution of the results. However, such deviations from the hardware seem to evade the noisy simulations in Qiskit, possibly because of the simplified noise in classical simulations. For example, while the Qiskit Aer simulator can create a noise model that is based on calibration data from a real system, these values are only the average error rates and does not account for instantaneous errors that can occur but do not meaningfully shift the average error rate. Thus, while the simulators can give an estimation of the expected results, they represent a best-case scenario and can miss the more complex and currently not well defined errors \cite{doi:10.1021/acs.chemrev.4c00870} that occur on real NISQ hardware.

The current NISQ computers typically has a limited scale of functional qubits ($\approx 100$) and a typical error rate of 1 error for about 1000 gate operations \cite{Preskill_2018, IBMQPUData, Chen2023}. According to the splitting of our circuit into the state preparation, Uhlmann process, and extraction of expectation values, we can look into the amount of gate operations needed to realize each block of the circuit. These results are summarized in Table. \ref{tab:nonOptimzedISACounts}, where we can see that the extraction of expectation values takes the least amount of total operations for either the $\langle \sigma_x \rangle $ or $\langle \sigma_x \rangle$ measurements. However, we find that the arbitrary state preparation scheme takes almost 2000 ISA gate operations on its own to create the desired purified state. Similarly, the full Uhlmann process which consists of the two $C\tilde{\mathcal{R}_{y}}$ gates also uses just over 300 ISA gate operations. We find then that the most immediate and naive construction of these circuits contain $\approx$ 2400 ISA gate operations in total, well above the average error rates of current NISQ systems. As shown in Fig.~\ref{fig:UhlmannResults}(a), the QPU results from the naive circuit cannot resolve the jumps of the spin-1 Uhlmann phase. We will implement circuit optimization in the following section and discuss an error-mitigation method in a later subsection.

\begin{table}
    \centering
    
    \begin{tabular}{|c|c|c|}
    \hline
        & spin-1/2 & spin-1 \\
        \hline
        State Preparation & 18 & 2090 \\ 
        \hline
        Uhlmann Evolution & 26 & 320 \\
        \hline
        Trace Estimation & 4 & 4 \\
        \hline
    \end{tabular}
    
    \caption{Average number of ISA gate operations when transpiled for IBM-Sherbrooke for the spin-1 and spin-$\frac{1}{2}$ systems, using the Qiskit synthesis and transpiler with optimization level 1. For the spin-1 case, we consider the naive circuit which uses the arbitrary state preparation rather than the optimized Qiskit state preparation. The average is found over the temperature range $ 0.01 \leq T <1$.}
    \label{tab:nonOptimzedISACounts}
    
\end{table}

    
    
    

    


\subsection{Optimization of the circuit}
As shown above, the circuit with the arbitrary state preparation generated from the Qiskit compiler is too large for the functional limits of
the IBM NISQ QPUs. 
In the following, we discuss tradeoffs between several customized circuit generation pipelines using unitary synthesis.

We first replace the arbitrary state preparation by the Qiskit's {\it StatePreparation} method based on exact synthesis using Quantum Shannon Decomposition~\cite{Shende_2006}. This reduces number of operations for the state preparation of the spin-1 circuit from about 2000 to about 100 ISA gates. With the substantially reduced gate count, the fidelity is also substantially improved, which will be discussed later. Importantly, the circuit size for the state preparation in the spin-1 circuit is no longer close to the NISQ limit after the optimization is implemented, and the prepared state is much closer to the desired purified state. Meanwhile, as the asymptotic circuit complexity of Qiskit's generic unitary synthesis is $O(4^n)$, we do not see substantial gate reduction when applying it to the second circuit component for the Uhlmann process.

To further reduce gate count, we use the BQSKit~\cite{osti_1785933} synthesis based compiler. While Qiskit uses an exact analytical decomposition method, BQSKit uses a suite of approximate synthesis algorithms~\cite{QSEARCH,QFAST,LEAP10.1145/3548693} combined with circuit partitioning~\cite{QGO} and an exploration of circuit permutations~\cite{PAM}. At the expense of compilation time, BQSKit can produce highly optimized circuits while allowing control over approximation level. Approximation is measured using the Hilbert-Schmidt distance between the target and generated unitary operations, and it limited by machine precision to roughly distances of $\varepsilon = 10^{-20}$ with a default value of $\varepsilon = 10^{-8}$. A discussion of the improvement and computational time when $\varepsilon$ varies will be presented later.

\begin{table}
    \centering

    \begin{tabular}{|c|c|c|}
    \hline
        U3 and CNOT & Qiskit synthesis & BQSKit synthesis \\
        \hline
         Arb State Prep & 170 U3, 196 CNOT & 174 U3, 63 CNOT \\
        \hline
         Qiskit State Prep & 15 U3, 17 CNOT  & 38 U3, 12 CNOT \\
         \hline
         Uhlmann Process &  64 U3, 36 CNOT & 12 U3, 4 CNOT \\
         \hline \hline
         Total ISA Gates & & \\
         \hline
         Arb State Prep & 2090 & 966\\
         \hline
         Qiskit State Prep & 114 & 159\\
         \hline
         Uhlmann Process & 320 & 59\\
         \hline
    \end{tabular}

    \caption{Average number of gate operations for different blocks of the the spin-1 circuit after the Qiskit and BQSKit optimizations. The average is found over the temperature range $ 0.01 \leq T <1$, and $\varepsilon=10^{-8}$ for BQSKit. The first four rows consider both the amount of three-angle unitary (U3) and CNOT gates for the most general unitary synthesis, and the lower three rows show the total number of ISA gates after transpiling for the IBM-Sherbrooke QPU.
    }
    \label{tab:generalGateNumbers}
    
\end{table}

The BQSKit optimization was preformed on the transpiled circuits given
by Qiskit with the qubit routing and selection handled, thus BQSKit
only handles the optimization of these circuit operations rather than
the full compilation problem. When using the default approximation distance, we find that BQSKit can achieve further
optimization that results in a substantial reduction in the amount of
three-angle unitary (U3) and CNOT gates used when compared to those
provided by Qiskit. These results can be seen in Table
\ref{tab:generalGateNumbers}, along with the direct reduction in the number of the ISA gates used for each block. We note that while we saw an increase in the total number of ISA and U3 gates when applying BQSKit to the Qiskit state preparation, we find an apparent trade off in that the number of CNOT gates is reduced as the U3 count is increased. A similar trend can also be seen with the arbitrary state preparation, suggesting that the optimizations found by BQSKit are of a potentially higher fidelity for NISQ systems as they trade the typically noisier two-qubit gates for an increase in the less noisy one-qubit gates.




    

Importantly, the optimization results in a circuit for the spin-1 system that now has a gate count well below the NISQ limit, thereby allowing a reliable simulation and extraction of the Uhlmann phase.
We ran the optimized spin-1 circuit through both the calibrated Aer simulator and the IBM-Sherbrooke QPU and show the results in Fig. \ref{fig:UhlmannResults}(b). Comparing these results to our previous results of the naive circuit in Fig. \ref{fig:UhlmannResults}(a), the simulation results are already improved after the optimization due to the substantially smaller gate count.

More importantly, the intermediate-temperature topological regime characterized by the quantized value of the Uhlmann phase now lands almost exclusively in the predicted region. Therefore, the optimized circuit of the spin-1 system reveals interesting finite-temperature topological phenomenon on real quantum hardware, which may be otherwise challenging for natural systems due to the requirement of satisfying the Uhlmann parallel-transport condition by controlling the system and environment in a coordinated fashion~\cite{OurDPUP}.
We note errors from the IBM NISQ systems still result in fluctuations of the data, but the quantized Uhlmann phase is now resolvable from the background noise.
Our results and analyses indicate that the spin-1 circuit with the Qiskit StatePreparation and BQSKit has reached optimization such that further tuning only creates minimal quantitative differences.




\subsection{Hardware upgrade}
Recently, IBM has started to retire its older, Eagle R1 based systems such as IBM-Sherbrooke used in Fig. \ref{fig:UhlmannResults}(b) in favor of their newer Heron R2 architecture. These newer systems promise improved 2-qubit gate infidelity of $9.63\times10^{-4}$ when compared to that of Eagle at $2.88 \times 10^{-3}$ \cite{AbuGhanem2025}. With part of our work being done on IBM-Sherbrooke, it was important to make sure that our results are portable between these different systems. Using the same circuits and BQSKit optimization workflow, we re-transpiled these circuits for the newer IBM-Kingston system which uses the Heron R2 architecture. It is important to note that this change in architecture comes with a change in the native gate sets. For the Eagle R1 systems this consists of ${ECR, RZ, X, SX}$ whereas for the Heron R2 systems the gate set is ${RZ, RX, CNOT, SX, X, RZZ}$, thus we cannot simply use the same transpiled circuits between different systems as they differ in their gate sets and therefore have different optimization procedures to be found by Qiskit and BQSKit. 

Using the same circuit-generation procedure as shown before, we re-transpiled, optimized with Qiskit and BQSKit, and ran these circuits on the newer Heron R2 based IBM-Kingston system. These results can be seen in Fig.~\ref{fig:UhlmannResults}(c) where we find that our results continue the same trend previously seen and can readily see the intermediate-temperature topological regime characterized by the quantized Uhlmann phase only at finite temperatures. We also note that the slight fluctuations that can be seen in the IBM-Sherbrooke data, particularly in the regions of what should be $\theta_{U} = 0$, are reduced even further and achieve a value that is much closer to zero, signaling the tangible improvements that have been made between these systems.

We also investigated the robustness of the result against the topology of qubit assignment on the QPU. When running the transpiliation on the circuits, Qiskit will find a selection of five qubits with suitable qubit-to-qubit connectivity needed for the spin-1 circuit operations while attempting to find the least noisy qubits available. This choice may be made differently each time due to the stochastic nature of the selection algorithm, thus by re-running the transpiler, we can assign the circuit to different selections of qubits on the QPU. After re-running the transpiler on the optimized circuits and ensuring that the qubit assignments differed, we ran these circuits on both IBM systems, Sherbrooke and Kingston, and found no qualitative difference of the measured results with respect to the topology of qubit assignment within each QPU. This further shows that the final optimized circuits are well below the NISQ limit, thereby removing any potential dependence on the topology of qubits executing the computation.

\subsection{Intermediate-temperature topological regime}
Optimizing the state preparation and Uhlmann process leads to the improvement from  Fig.~\ref{fig:UhlmannResults}(a) to (b), and upgrading of the hardware further enhance the quality of data from Fig.~\ref{fig:UhlmannResults}(b) to (c). The intermediate-temperature topological regime characterized by a quantized Uhlmann phase sandwiched between trivial low- and high-temperature regimes for the spin-1 system is now clearly visible from the IBM QPU. The results show how temperature affects the density matrices and through the purification and parallel transport, the underlying topology reflected by the Uhlmann holonomy is changing accordingly.

As illustrated in Appendix~\ref{app:spin_half} for typical spin-1/2 systems, the topological regime with a finite Uhlmann phase is an extension from the zero-temperature limit, making the topological phase transition more conventional as the system transits to a topologically trivial regime at high temperatures. The spin-1 system illustrated in Fig.~\ref{fig:UhlmannResults}(c) thus gives a concrete example of the complex interplay between topology and temperature in quantum systems, which is made possible by the versatile quantum computer for manipulating the purified state consistently with the Uhlmann process.

\section{Discussion}\label{sec:discussion}

\begin{figure}[t]
    \centering
    \includegraphics[width=0.8\linewidth]{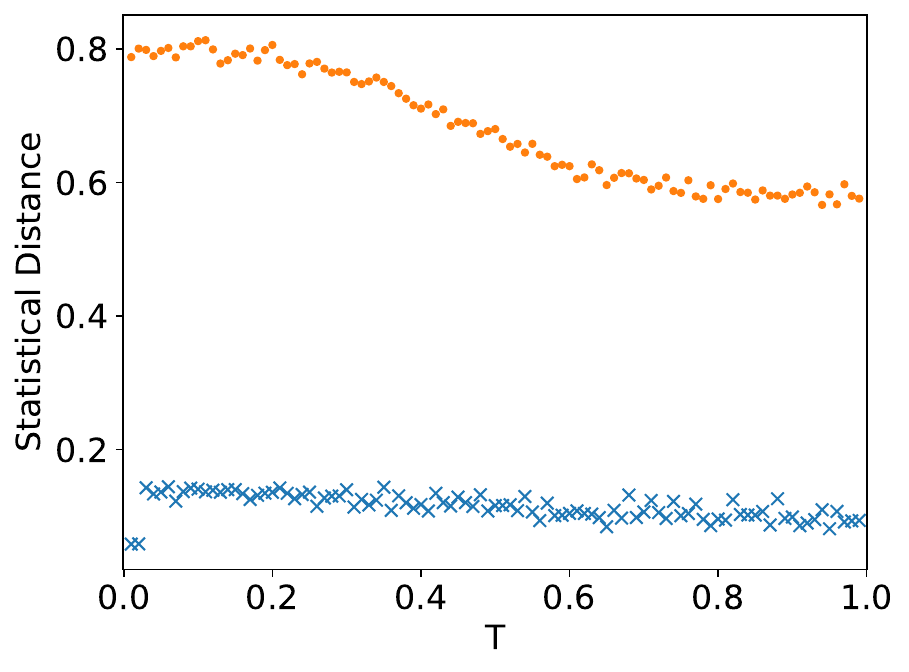}
    \caption{Statistical distance from the exact state for the arbitrary state preparation (dot symbols) and the Qiskit state preparation (cross symbols) for the spin-1 system. Both state preparation circuits were run using the transpiled circuits on the Aer simulator using calibration data from IBM-Sherbrooke. 
    }
    \label{fig:StatisicalDistanceFig}
\end{figure}

\subsection{Measures of fidelity for state preparation}
The Qiskit state preparation cuts the gate count by one order from the arbitrary state preparation. We quantify the accuracy of the state preparation through the total variation distance \cite{JACOBS2018173}, also called the statistical distance, 
defined as 
\begin{equation}
    \Delta S = \frac{1}{2} \sum_{x}|P(x)_{a} - P(x)_{b}|,
\end{equation}
where $P(x)_{a,b}$ are the two probability distributions being compared.
Measuring with respect to the desired purified state, the state preparation performs better when $\Delta S $ is small.
We calculated the distance with respect to the target state over the whole temperature range of interest for both the arbitrary and Qiskit state preparations. 

These results are shown in Fig. \ref{fig:StatisicalDistanceFig}. The state preparation preformed by the Qiskit method results in a state that is much closer to the desired state, although not exactly perfect due to the noise from the hardware, whereas the arbitrary state preparation resulted in a much larger difference from the exact state. Interestingly, we find that for both forms of state preparation the statistical distance from the target state decreases as the temperature is increased. This is understandable since the distribution of a high-temperature state becomes uniform, which better accommodates the random noise.




\subsection{BQSKit approximation distance}
Investigating the effect of BQSKit's approximation of the spin-1 circuit, we varied the approximation distance $\varepsilon$ through four different values from a more coarse-grained ($\varepsilon = 10^{-3}$) to a fine-grained approximation ($\varepsilon = 10^{-15}$). The total number of ISA gate operations after these different approximations are shown in Table. \ref{tab:bqskitApproxGateCounts}. Here we see that increasing the apparent precision of the approximation by moving from $\varepsilon = 10^{-3}$ to $\varepsilon = 10^{-10}$ does not increase the average number of ISA gates by a substantial amount, increasing by $\approx 30$ gates at most. Moving to the fine-grained optimization at $\varepsilon = 10^{-15}$, we see that the ISA gate count increases substantially and comes close the the NISQ limit. Moreover, the gate count with $\varepsilon = 10^{-15}$ even exceeds the number without BQSKit, defying the purpose of optimization. Therefore, our analysis suggests that moderately coarse-grained optimizations produce circuits more suited for use on current NISQ systems as opposed to the computationally expensive fine-grained approximation. These BQSKit approximations were preformed on the UC Merced Pinnacles cluster consisting of Intel Xenon 6330 based nodes.

\begin{figure}
    \centering
    \includegraphics[width=0.85\linewidth]{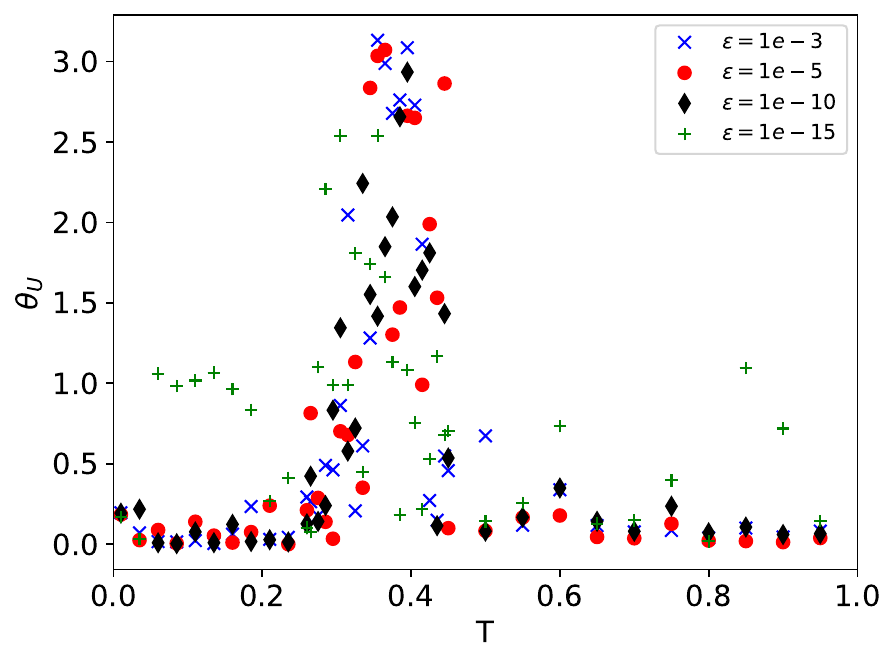}
    \caption{Uhlmann-phase results from IBM-Sherbooke for different approximation distances $\varepsilon=10^{-3}, 10^{-5}, 10^{-10}, 10^{-15}$ from the target unitary operations used in the BQSKit synthesis. These circuits use the same spin-1 circuits with the Qiskit StatePreparation, Uhlmann process, and measurement.}
    \label{fig:bqskitSynthesisUhlmannResults}
\end{figure}

Following the discussion on the BQSKit approximation, we ran the circuits corresponding to selected approximation distances to see if they might lead to further improvement or refinement of the results. Interestingly, we find that running these circuits on IBM-Sherbrooke produces results that all distinctly show the intermediate-temperature topological regime and lie within the same temperature range even with the expected fluctuations due to the inherent noise of the QPU, as shown in  Fig. \ref{fig:bqskitSynthesisUhlmannResults}. However, the $\varepsilon=10^{-15}$ results suffer stronger fluctuations due to its large circuit size. Our analyses thus suggest that it is appropriate to use the default approximation distance ($\varepsilon=10^{-8}$) for BQSKit optimization for the current IBM QPUs. Moreover, given that we saw a large jump in the computational time taken to produce these fine-grained approximations, going from $\approx 3.5hrs$ to $\approx 12.5 hrs$, it further shows that these finer approximations are not the best use of computational resources for circuits to be run on current NISQ hardware.

\begin{table}[t]
    \centering
    \begin{tabular}{|c|c|c|}
    \hline
        &  Number of ISA Gates & Computational time \\ \hline
         $\varepsilon = 10^{-3}$ &  192  & $\approx$ 4hrs \\ \hline
         $\varepsilon = 10^{-5}$ &  195  & $\approx$ 5 hrs\\ \hline
         $\varepsilon = 10^{-8}$ &  221 & $\approx$ 3.5 hrs\\ \hline
         $\varepsilon = 10^{-10}$ &  227 & $\approx$ 5 hrs\\ \hline
         $\varepsilon = 10^{-15}$ &  926 & $\approx$ 12.5 hrs\\ \hline
    \end{tabular}
    \caption{Average number of ISA gate operations for the optimized spin-1 circuit using BQSKit unitary synthesis at different approximation distances $\varepsilon$ from the target unitary operation. The average is found over the temperature range $ 0.01 \leq T <1$ for all of the circuits. The circuit includes the Qiskit StatePreparation, the Uhlmann Process, and the measurement. The computational time is the amount of time for the BQSKit optimization on all of the circuits used over the same temperature range.
    }
    \label{tab:bqskitApproxGateCounts}
\end{table}

\subsection{Dynamic decoupling}
Another possible scheme to reduce the accumulation of errors is the error mitigation technique of dynamical decoupling (DD) \cite{PhysRevA.58.2733}. This error mitigation method seeks to eliminate errors that come from a comparatively long time where the qubit is not being manipulated, increasing the chance for the qubit to decohere into an unwanted state. This dechoerence is caused by an unwanted form of coupling between the qubit and the external system, that despite best efforts in construction and isolating the QPU, might still persist during computations. By applying a series of pulses that result in the identity, i.e., $\hat{U}_{1} \hat{U}_{2} \hat{U}_{3} \cdots \hat{U}_{n} = \hat{I}$, the qubit can be forced to maintain its coherence as its state is moved around the Bloch sphere, reducing the coupling between the qubits and environment, and thereby reducing the chance for errors from decoherence.

\begin{figure}[t]
    \centering
    \includegraphics[width=0.9\columnwidth]{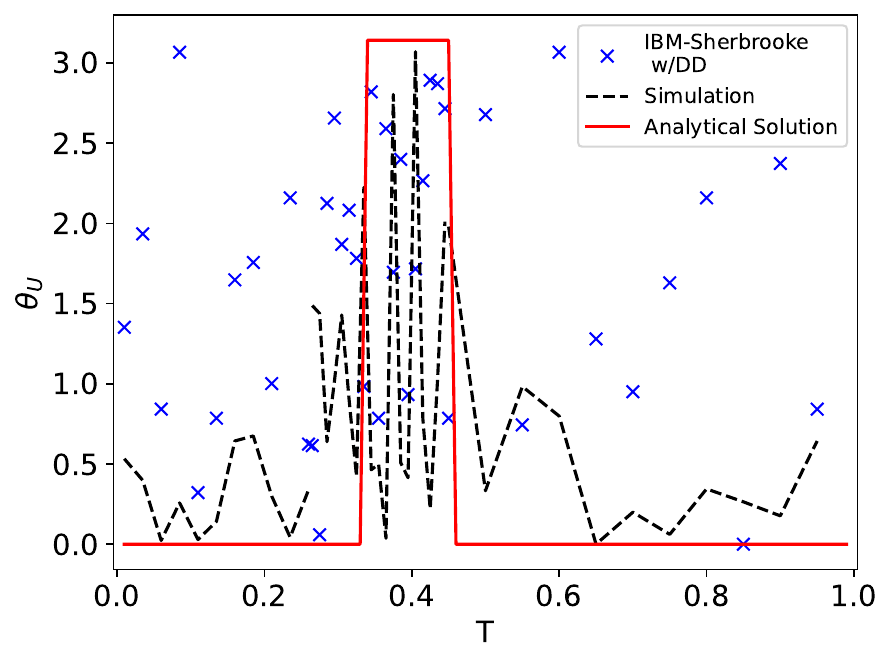}
    \caption{Uhlmann phase for the spin-1 circuits with dynamical decoupling (DD) but arbitrary state preparation and Uhlmann process before BQSkit optimization. The solid line, dashed line, and cross symbols show the analytic result, simulations with noise, and QPU result. There is no noticeable improvement over the results without DD shown in Fig.~\ref{fig:UhlmannResults}(a).}
    \label{fig:uhlmannWDD}
\end{figure}

For the Uhlmann phase circuit, we successfully implemented this error reduction scheme in Qiskit and on IBM's QPUs using a well known gate sequence called XY4 \cite{MAUDSLEY1986488}.  This sequence consists of the X and Y gates applied in pairs, $XYXY|\psi\rangle$, which is equivalent to the identity. After running this augmented circuit without optimization on the IBM Sherbrooke QPU, we saw no tangible improvement on our results that would bring them inline with those found using the simulator, as shown in Fig. \ref{fig:uhlmannWDD}. Therefore, adding the DD to the circuit without optimization shows no marked improvement over the results without the DD. While the DD scheme is a proven method for reducing potential errors in superconducting qubits \cite{PhysRevApplied.20.064027}, from our null-results we find that our source of error is most likely not caused by decoherence of the qubits. We attribute this lack of improvement due to the fact that while dynamical decoupling has been shown to reduce dechoerence errors on average, the effectiveness of the error mitigation crucially depends on the type of noise that is present in the system \cite{PhysRevApplied.22.054074}. When working with superconducting qubits, pinpointing the exact form of this error can be a difficult job even with direct access to the system itself \cite{PhysRevA.58.2733}. Indeed then, cutting down the gate counts via optimization of state preparation and Uhlmann process is shown to be the proper means for extracting the Uhlmann phase on IBM's NISQ computers.

\subsection{Comment on quantum phase estimation}
Quantum phase estimation~\cite{ kitaev1995quantummeasurementsabelianstabilizer, Nielsen_Chuang_2010} extracts the phase of the eigenvalues $\exp(i\theta)$ of a unitary operator $U$ if the eigenstate $|\psi_U\rangle$ is given. However, the Uhlmann phase is not the phase of the eigenvalue of the Uhlmann process on the purified state. Instead, the Uhlmann phase is the phase from the transition amplitude $\langle W(\tau)|W(0)\rangle$ between the initial and final amplitudes of the corresponding density matrices. Moreover, the purified state is not an eigenfunction of the unitary operator representing the Uhlmann process. Therefore, the Uhlmann phase cannot be extracted by quantum phase estimation in a direct fashion. Instead, we follow Ref.~\cite{Viyuela2018} and apply the generalized DQC1 scheme to extract the Uhlmann phase by coupling the mixed state with a probe qubit.

\subsection{Other applications}
The construction of the quantum circuit of Fig.~\ref{fig:circuit} can be applied to higher spin-$j$ systems by using $2j$ qubits to realize the system plus $2j$ qubits for the ancilla, in addition to the probe qubit. However, one has to project out the states outside the spin-$j$ space, as illustrated by the spin-1 system studied here. For the state preparation, it still follows the purification of the density matrix to prepare the corresponding purified state for the system plus ancilla. The unitary operators $U_s$ and $U_a$ for generating the Uhlmann process are realized by the corresponding rotation operators. Finally, the probe qubit controls the rotation operators for generating the Uhlmann process and its expectation values $\langle\sigma_{x,y}\rangle$ reveals the Uhlmann phase. The complexity of state preparation and Uhlmann process will increase rapidly with $j$. Interestingly, one may view the realization and measurement of the Uhlmann phase of higher spin-$j$ system as a benchmark for NISQ hardware when the gate count approaches the error budget of the available quantum computer.

We mention that there are other mixed-state geometric phases in the literature besides the Uhlmann phase by imposing different parallel-transport conditions or underlying geometric structures. One example is the interferometric geometric phase (IGP)~\cite{PhysRevLett.85.2845,PhysRevA.67.020101,PhysRevA.70.052109,Faria_2003,Chaturvedi2004,PhysRevLett.93.080405,Kwek_2006}, which becomes the thermal weighted sum of the Berry phases of individual states in certain cases~\cite{PhysRevLett.85.2845,PhysRevA.68.022106}. For a unitary process $U_{IGP}(t)$ compatible with the IGP parallel-transport condition, the IGP is given by $\theta_{IGP}(t)=arg[Tr(\rho(t=0)U_{IGP}(t)]$ when a set of parallel-transport conditions are satisfied. One can see that the IGP also has a form suitable for the modified DQC1 scheme that we implemented here for the Uhlmann phase. Interestingly, the IGP also exhibits quantized jumps at finite temperatures~\cite{PhysRevB.107.165415,Wang25}, but the former typically shows less features than the latter. We also caution that while the Uhlmann phase represents the Uhlmann holonomy of the bundle of density matrices, the IGP has not been associated directly with a topological object.

\section{Conclusion}\label{sec:conclusion}
We have demonstrated the intermediate-temperature topological regime characterized by the Uhlmann phase for the spin-1 system on IBM quantum computers. While the straightforward construction of the quantum circuit shows promising results from the noisy simulations, the real QPU results have high levels of fluctuations and could not resolve the discrete values of the Uhlmann phase. Nevertheless, optimizations in the state preparation and Uhlmann process through Qiskit and BQSKit reduces the circuit size dramatically, making the signals from the QPU visible. The timely upgrade of the QPU further reduces the noise. The latest QPU results now follow closely with the analytic prediction. The framework introduces a class of physical problems that can be studied by NISQ computers. On the other hand, the Uhlmann phase of higher spin-$j$ systems may serve as a benchmark for checking NISQ hardware and software capabilities since the qubit count and circuit depth increases with $j$ substantially.

\begin{acknowledgments}
C. M. and C. C. C are supported by the NSF (No. PHY-2310656) and DOE (No. DE-SC0025809). We acknowledge the use of IBM Quantum Credits for this work. Computational time supporting this research was provided by the UC Merced Pinnacles cluster (NSF MRI, No. 2019144). The views expressed are those of the authors, and do not reflect the official policy or position of IBM or the IBM Quantum team.
\end{acknowledgments}

\appendix
\section{Uhlmann phase for spin-1/2 systems}\label{app:spin_half}
For the parallel-transport of a spin-1/2 system along a longitude in the parameter space, the rotation angles for the system and ancilla are $\theta_{s} = \int_{0}^{t}\theta^{\prime}(t) dt^{\prime} = 2\pi$ and $\theta_{a} = \eta\theta_{s} = 2\eta \pi$, respectively. When taking $\omega_{0} = 1$, the time-evolution operators for the system and ancilla in order to generate the Uhlmann process are given by
\begin{eqnarray}
    U_{s} &=& 
    \begin{pmatrix}
        \cos(\pi) & -\sin(\pi) \\
        \sin(\pi) & \cos(\pi)
    \end{pmatrix}
    =     \begin{pmatrix}
        -1 & 0 \\
        0 & -1
    \end{pmatrix}, \nonumber \\
    U_{a} &=& 
    \begin{pmatrix}
        \cos(\pi ~ sech(\frac{\beta}{2})) & -\sin(\pi ~ sech(\frac{\beta}{2})) \\
        \sin(\pi ~ sech(\frac{\beta}{2})) & \cos(\pi ~ sech(\frac{\beta}{2}))
    \end{pmatrix}.
\end{eqnarray}

\begin{figure}[t]
    \centering
    \includegraphics[width=0.9\linewidth]{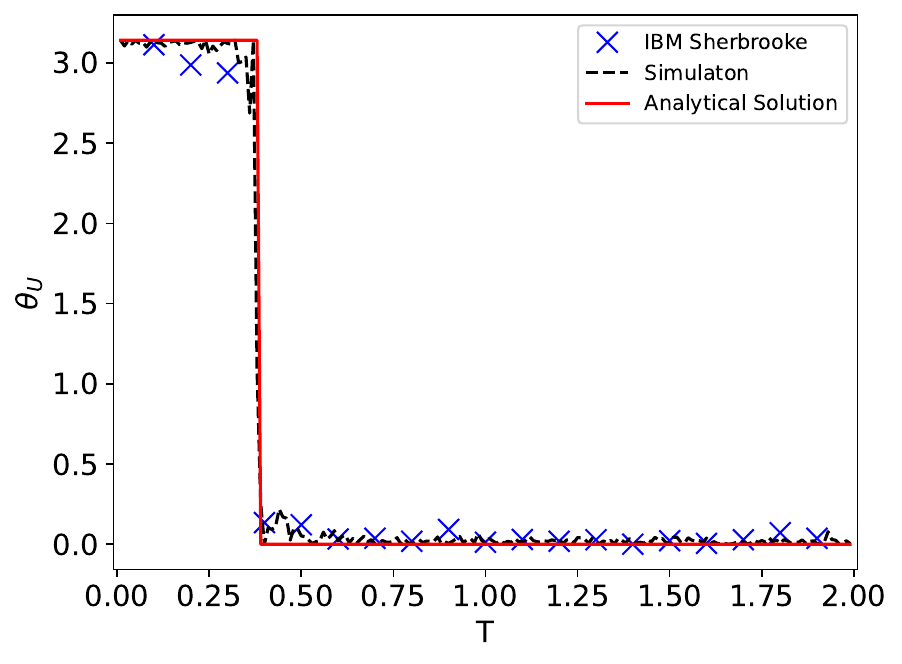}
    \caption{Uhlmann phase as a function of temperature for the exact solution (solid line), simulation (dashed line), and IBM-Sherbrooke QPU results (cross symbols) for the spin-$\frac{1}{2}$ case. The jump from $\theta_{u} = \pi$ to $\theta_{u} = 0$ signifies a topological phase transition.}
    \label{fig:spin12-results}
\end{figure}

Fig. \ref{fig:spin12-results} shows the results for the Uhlmann phase calculated with the Qiskit simulator and running on IBM's Sherbrooke QPU. The analytic solution for the Uhlmann phase is shown in the red solid curve, and we can see that both the simulation and QPU results follow this curve well even with potential errors caused by the noise present in NISQ hardware. However, the transition is from the low-temperature topological regime to the high-temperature regime for the spin-1/2 case. Ref.~\cite{OurPRA21} considers more exotic spin-1/2 cases with higher internal windings to generate the intermediate-temperature topological regime. However, the spin-1 case naturally exhibits such a phenomenon as shown in the main text.

We further note that the $\langle \sigma_{y} \rangle$ averaged across all temperatures is found to be $\langle \sigma_{y} \rangle = 0.035$ on the IBM Sherbrooke QPU, which is an improvement of $\approx 280\%$ over the corresponding value of $ 0.098 \pm 0.014$ found in Ref.~\cite{Viyuela2018} for the now retired ibmqx2 system, showing the overall improvement that IBM's quantum computers have attained within the past few years. 
On the other hand, the spin-1 case also sees substantial improvement from the hardware upgrade shown in Fig.~\ref{fig:UhlmannResults}(b) and (c).

%

\end{document}